# Using metal-organic frameworks to confine liquid samples for nanoscale NV-NMR

Kristina S. Liu[1§], Xiaoxin Ma[1§], Roberto Rizzato[1], A. Lisa Semrau[1], Alex Henning[2], Ian D. Sharp[2], Roland A. Fischer[1], Dominik B. Bucher[1*]

[1]Department of Chemistry, Technical University of Munich, 85748 Garching, Germany

[2]Walter Schottky Institute and Physics Department, Technical University of Munich, 85748 Garching, Germany

ABSTRACT

Atomic-scale magnetic field sensors based on nitrogen vacancy (NV) defects in diamonds are an exciting platform for nanoscale nuclear magnetic resonance (NMR) spectroscopy. The detection of NMR signals from a few zeptoliters to single molecules or even single nuclear spins has been demonstrated using NV-centers close to the diamond surface. However, fast molecular diffusion of sample molecules in and out of nanoscale detection volumes impedes their detection and limits current experiments to solid-state or highly viscous samples. Here, we show that restricting diffusion by confinement enables nanoscale NMR spectroscopy of liquid samples. Our approach uses metal-organic frameworks (MOF) with angstrom-sized pores on a diamond chip to trap sample molecules near the NV-centers. This enables the detection of NMR signals from a liquid sample, which would not be detectable without confinement. These results set the route for nanoscale liquid-phase NMR with high spectral resolution.

TEXT

INTRODUCTION

Nuclear magnetic resonance (NMR) spectroscopy is a powerful chemical characterization tool widely used in disciplines ranging from materials to biomedical sciences. However, it is limited by low sensitivity, typically requiring a substantial amount of material for analysis[1]. One promising way to overcome this limitation is via the use of nanoscale spin defects in solids, which can be brought in close contact with the sample[2–6]. The most promising system is the nitrogen vacancy (NV) center in diamond, which has been used to detect NMR signals from a few zeptoliters of material, single molecules or even nuclear spins under ambient conditions[7,8]. The NV-center (Fig. 1 A) is a spin defect in the lattice of a diamond. The spin quantum states of such defect centers can be optically prepared with green laser light, manipulated with microwave pulses and optically read-out by their spin-dependent photoluminescence (PL).[9] Dynamic decoupling pulse sequences (such as XY8-$N$, where $N$ is the number of $\pi$ pulses ) can be used as a narrowband detection scheme for oscillating magnetic fields at frequency $f = 1/(2\tau)$, where $\tau$ is the temporal spacing between the microwave $\pi$ pulses (Fig. 1 B). Sweeping the time $t_{corr}$ between two XY8-$N$ blocks correlates the magnetic signals from the nuclear spins, which can be detected as oscillations in the PL readout as a function of $t_{corr}$ and resembles the free induction decay in traditional NMR spectroscopy (Fig. 1

B)[10–12]. The detection volume of the NV-center from which NMR signals are detected corresponds to a hemisphere whose radius is roughly given by the depth of the defect below the surface of the diamond (e.g., 5 to 10 nm)[2,4]. At this sample volume – in the zeptoliter range – thermal polarization of the nuclear spins is negligible and statistical polarization from incompletely cancelled spin magnetization dominates[12,13]. For that reason, the NMR signal strength is independent of the applied magnetic field $B_0$ and the detection does not require any active nuclear sample spin manipulation[3,12,14,15].

Despite the advantages described above, nanoscale NMR comes with its own challenges. The main drawback is that the time scale of diffusion of liquid samples across nanoscale detection volumes limits the interaction of the samples with the NV-sensor, which leads to broadened signals for viscous samples and to undetectable signals for low viscosity liquid samples[7,12,15–17]. Within sensing volumes, liquid-state molecules can diffuse in and out because of Brownian motion (Fig. 1 C, left). For illustrative purposes, one can visualize the effect that molecular diffusion has on the nuclear spin signal by simulating a random walk of a molecule at a certain distance from the NV center, with its molecular mobility physically restricted by the presence of the diamond surface. Then, following the approach in Pham et al.[18], the coupling factor $c = \frac{u_z^2(1-u_z^2)}{r^6}$ can be calculated, where $r$ is the length of the vector connecting the NV center and the molecule and $u_z$ contains the angular terms in spherical coordinates. This coupling factor estimates how strongly the target nucleus interacts with the NV sensor at a certain time point within the experimental timescale. The decay of this value over time, caused by the average distance of the molecules from the NV center over time, ultimately results in the expected NV-NMR linewidth.

Here, two different exemplary conditions for molecular mobility can be envisioned, as represented in Figure 1 C. With a diffusion coefficient of $D = 5\times10^{-10}$ m$^2$/s (ten times lower than water), the probability of a molecule free to move in the vicinity of the diamond remaining within the sensing volume of the NV center drops off rapidly in time (Fig. 1 C, left). This has a dramatic impact on the NV-based NMR sensing, yielding signals that are too broad to be detected experimentally. For a slow diffusion coefficient of $D = 5\times10^{-15}$ m$^2$/s (Fig. 1 C, right) a much narrower linewidth is expected, allowing for a detectable signal. Consequently, the restriction of diffusion, while keeping the sample in the liquid state to avoid dipolar broadening, is crucially important for NV-NMR. Prior theoretical work raised a compelling possibility for counteracting the effects of diffusion by confinement of liquid samples in small volumes, thereby enabling this major impediment to nanoscale NMR to be overcome[17].

An intriguing system for achieving such confinement is based on metal-organic frameworks (MOFs), which are porous materials with high surface areas[19,20]. Furthermore, the CoRE MOF database contains thousands of different MOFs with various physical and chemical properties, including different and precisely defined pore sizes[21]. This powerful capability enables specific MOFs to be rapidly identified that have suitable pore sizes and flexibility to efficiently impede diffusion by trapping small sample molecules[21]. In the present work, we selected the MOF UiO-66, which has been computationally identified for capturing $^{31}$P containing small molecules, resulting in slow diffusion coefficients[22]. This material can also be grown as a surface-anchored MOF, or "SURMOF", with well-ordered nanocrystals, achieving film thicknesses of less than 100 nm[23–27]. When grown on a NV-diamond, this UiO-66 SURMOF can realize the confinement of liquid-state nuclei within the nanoscale detection volumes of the NV-quantum sensors. Therefore here, we modify the diamond surface with thin films of UiO-66, which enabled us to observe $^{31}$P

NMR signals from the small phosphorous-containing molecule trimethyl phosphate (TMP). The size of TMP is comparable to those of other molecules that have been described by Agrawal et al.[22] as having very low diffusion coefficients in UiO-66, but with low toxicity. This proof-of-concept study can be generalized to the detection of a wide array of small molecules and sets the route to the detection of liquid-state samples using nanoscale NMR.

RESULTS

**Functionalizing diamond with UiO-66 SURMOF.** In order to restrict diffusion within the NV-detection volume, we functionalized the diamond surface with a thin layer (SURMOF) of UiO-66, which is a MOF with $Zr_6O_4(OH)_4$ nodes linked by 1,4-benzene-dicarboxylate (bdc).

The UiO-66 ($Zr_6O_4(OH)_4(bdc)_6$) was formed via coordination modulation-controlled step-by-step liquid-phase growth (Fig. 2 A)[25]. We chose to facilitate adhesion of the SURMOF by coating the diamond surface with a 1 nm thick layer of alumina ($Al_2O_3$) via atomic layer deposition (ALD)[28,29]. The alumina film is hydrophilic, as indicated by the relatively low static water contact angle (SWCA) of 18° ± 1.1° (Fig. 2 B), whereas the SURMOF surface is hydrophobic with an SWCA of 99° ± 3.6°. This is expected because the bdc linkers are poorly soluble in water (Fig. 2 B). Following 80 cycles of the liquid-phase MOF deposition yields a discontinuous film showing miniscule cracks on a micrometer scale and with an average thickness of ~85 nm (Fig. 2C). This is consistent with previous results from growth on a silicon dioxide surface[25]. The SURMOF on the diamond chip was then characterized by Raman spectroscopy, revealing known UiO-66 peaks in addition to a large diamond peak (Fig. 2 D)[30]. From these results, successful formation of a thin film of UiO-66 is confirmed.

**Characterization of SURMOF stability in and trapping of TMP within MOF pores**

Comparison of X-ray diffraction data obtained before and after soaking UiO-66 in trimethyl phosphate (TMP) confirms the stability of the MOF in solution (Fig. 2 E). This indicates that the SURMOF remains intact during the process applied to fill the MOF pores with TMP, which was accomplished via soaking in TMP overnight. TMP is soluble in $H_2O$, which we take advantage of to verify that the NV-detected NMR signal comes from within the pores using elemental analysis, as follows (Fig. 2 F). As expected, within the detection limit, the pure UiO-66 MOF powder contains no phosphorous (Fig. 2 F, condition A). After soaking the MOF in TMP for 2 days and filtering directly afterwards, the P content is ~7.6 wt. % (condition B), which decreases to ~3.2 wt. % after careful washing in $H_2O$ before filtering (condition C). The decrease of phosphorous content is likely due to removal of TMP from the surface of the MOF. This result suggests that the remaining phosphorous content is within the pores, which verifies the absorption of TMP by the MOF. Quantification of the elemental analysis indicates a molar ratio of 1:2 of UiO-66 MOF to TMP (see Supplementary Note 1: Elemental analysis of TMP and UiO-66 MOF powder). Pore volumes for UiO-66 have been reported to be ~ 0.5 $cm^3/g$[31–33] (see Supplementary Note 2: Estimation of pore filling from elemental analysis). From this we can estimate that the MOF contains approximately $1.9 \times 10^{24}$ pores/mol of MOF and that ~ 64% of these available pores are filled, meaning that TMP is well-absorbed by the MOF.

**Detecting NV-NMR-signals from small molecules trapped within SURMOF pores.** After functionalization, NV-NMR at 174 mT was performed on a shallowly implanted diamond (see Supplementary Note 3: Materials and Methods and Supplementary Note 4: Experimental Conditions) to detect $^{31}P$ NMR signal from neat TMP. We observe that the $Al_2O_3$-terminated diamond soaked in TMP shows no detectable $^{31}P$ signal (Fig. 3 A, left). Freely diffusing TMP molecules with Brownian motion on that order would produce an undetectable signal due to extremely broadened lines (Fig. 1 B).

In contrast to the fast diffusion of free TMP, slow diffusion due to confinement or adsorption within the pores is expected[22]. UiO-66 has two different types of pores, one with a size of 11 Å, called the primary pore, and a smaller one with a size of 7.2 Å, called the secondary pore. Each primary pore is connected to a secondary pore through a window of diameter 4.5 Å (Fig. 3 B). Due to the low molecular stiffness of UiO-66, physical hopping between pores is greatly slowed down for molecules that are larger than the window, which consequently strongly reduces the diffusivity[22]. Upon growing SURMOF on the oxide surface and filling the pores with TMP, we observe a narrow NV-NMR signal at ~3 MHz, which fits with the theoretical Larmor frequency at this magnetic field (Fig. 3 A, right). After rinsing and soaking the MOF-terminated diamond in $H_2O$ and then drying with nitrogen, the $^{31}P$ signal is still present (see Supplementary Note 5: Measuring $^{31}P$ signal from dried MOF). This further indicates that the $^{31}P$ signal arises from TMP within the pores since TMP is soluble in water and non-incorporated molecules would be readily cleaned from the surface of the MOF. Due to limited SNR, we are unable to completely resolve the linewidth (the spectrum shown has been zero-filled). However, we can claim an upper limit of ~ 3 kHz which is set by the specific $t_{\text{corr}}$ of the correlation measurement[34]. Importantly, this $^{31}P$ signal and associated linewidth is obtained exclusively from the MOF-terminated diamond, whereas no signal is detectable from freely diffusing TMP. We note that the data for both the MOF- and $Al_2O_3$-terminated surfaces soaked in TMP, as shown in Figure 3 A, are normalized to the reference signal (see Supplementary Note 6: Signal normalization using a 1 MHz reference signal).

The number of molecules within the detection volume can also be quantified by measuring the spin noise($B_{rms}^2$) using this 1 MHz reference signal. The calibrated strength of the $^{31}$P signal from TMP is determined to be ~ 920 nT$^2$ (Fig. 3 C). From this, we can estimate a density of TMP molecules trapped within the MOF pores using the equation[35]:

$$B_{rms}^2 = \frac{5\pi}{96}\left(\frac{\mu_0 h \gamma_{nuc}}{4\pi}\right)^2 \sigma \left[\frac{1}{d_{NV}^3} - \frac{1}{(d_{NV}+h_{spins})^3}\right], \quad (1)$$

where $\mu_o = 4\pi \times 10^7$ m·T/A is the vacuum permeability, h = 6.626 · 10$^{34}$ J·s is the Planck constant, $\gamma$ is the nuclear gyromagnetic ratio in MHz/T (17.235 MHz/T for $^{31}$P), $\varrho$ is the nuclear spin density, $h_{spins}$ is the thickness of the layer with nuclear spin, and $d_{NV}$ the depth of the NVs. Using the 85 nm average height of the MOF layer as the thickness and the simulated average NV-center depth of 4.5 nm[28,36] with an additional 1 nm from the Al$_2$O$_3$ deposited on the surface, we estimate that ~ 700 TMP molecules are trapped in each (10nm)$^3$. This calibration result is comparable to 530/(10nm)$^3$, the density of molecules calculated from the molar ratio determined from elemental analysis (see Supplementary Note 7: Calculation of TMP density within UiO-66 SURMOF pores from Elemental Analysis).

CONCLUSION

In summary, we show that restricting diffusion by confinement enables nanoscale NMR spectroscopy of liquid samples. Our approach uses MOFs with angstrom-sized pores on the surface of a NV-diamond chip to trap sample molecules near the NV-centers. This enables us to detect NMR signals from a liquid sample, which would not be detectable without confinement. We obtain the $^{31}$P signal from TMP trapped within the pores of UiO-66 grown as a thin layer on the diamond. Our results pave the way towards nanoscale liquid-state NMR with high spectral resolution. A great advantage of MOFs is that due to their reticular synthesis they can be tailor-made to have various physical and chemical properties, including pores of desired sizes, chemical compositions that do not interfere with molecules from the solution, and stabilities under different environments[37,38]. This method is therefore broadly usable for enrichment, colocalization and trapping of sample molecules near the NV-center for molecular analysis on a single molecule level[39] or NV-based hyperpolarization[40–42].

FIGURES

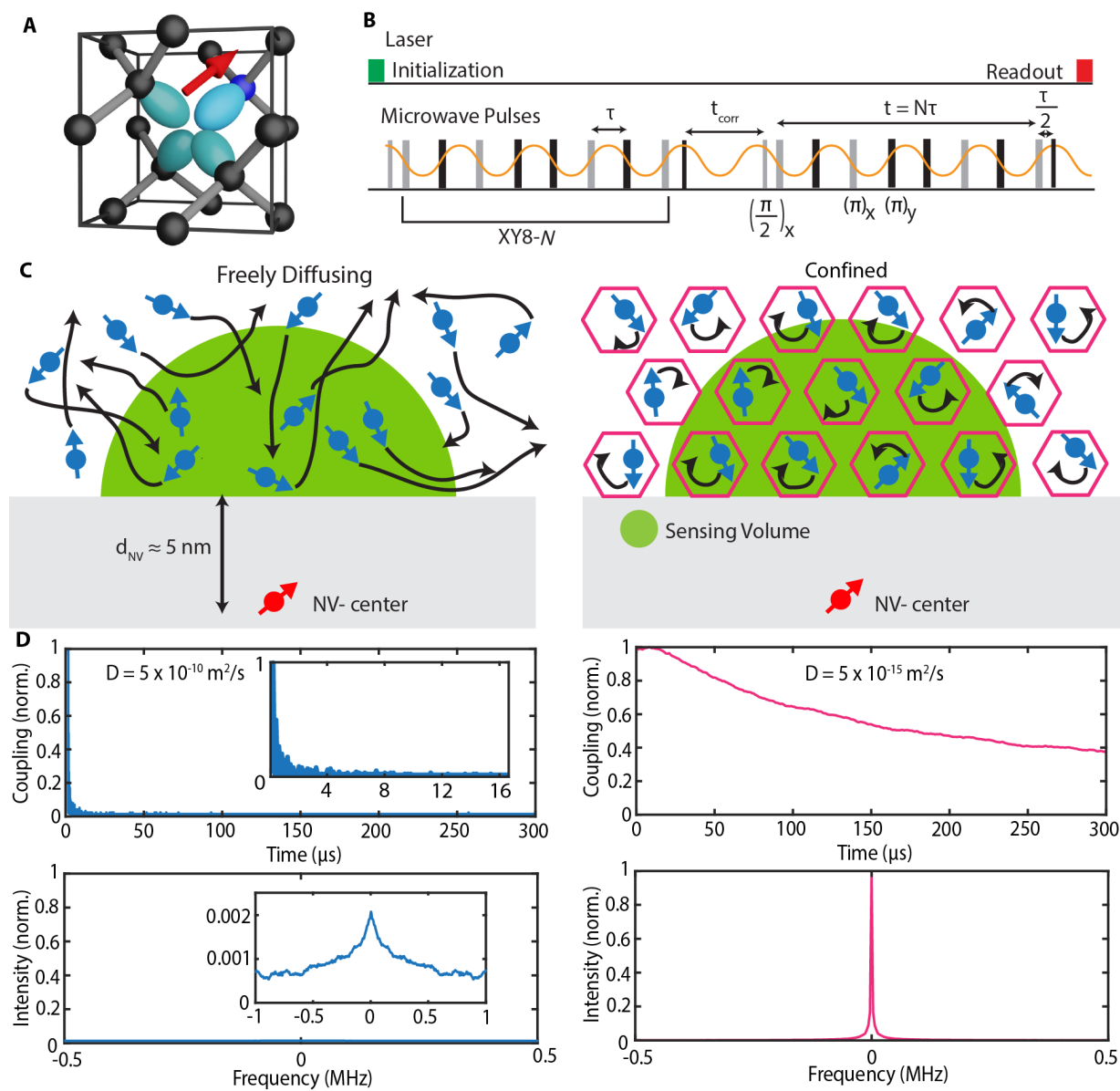

**Figure 1. Confinement of liquid samples for nanoscale NV-NMR spectroscopy. A)** Scheme of nitrogen vacancy (NV) center within the diamond lattice. The carbon atoms are shown in black,

the nitrogen atom in blue, the vacancy in the middle, and one potential orientation is indicated with a red arrow. **B)** Correlation spectroscopy pulse sequence for NMR signal detection. **C)** Schematic of nanoscale NV-NMR with liquid samples. Freely diffusing (blue) nuclei enter and leave the sensing volume (green) of the NV-centers while those confined (magenta) stay within the sensing volume. **D)** Fast diffusion limits the interaction time with the NV-center and causes a fast decay of the NMR signal (left), leading to severely broadened lines that are typically not detectable. Confinement of the spins (right) results in a long-lived NMR signal, enables its detection and opens a way to high spectral resolution nanoscale NV-NMR spectroscopy.

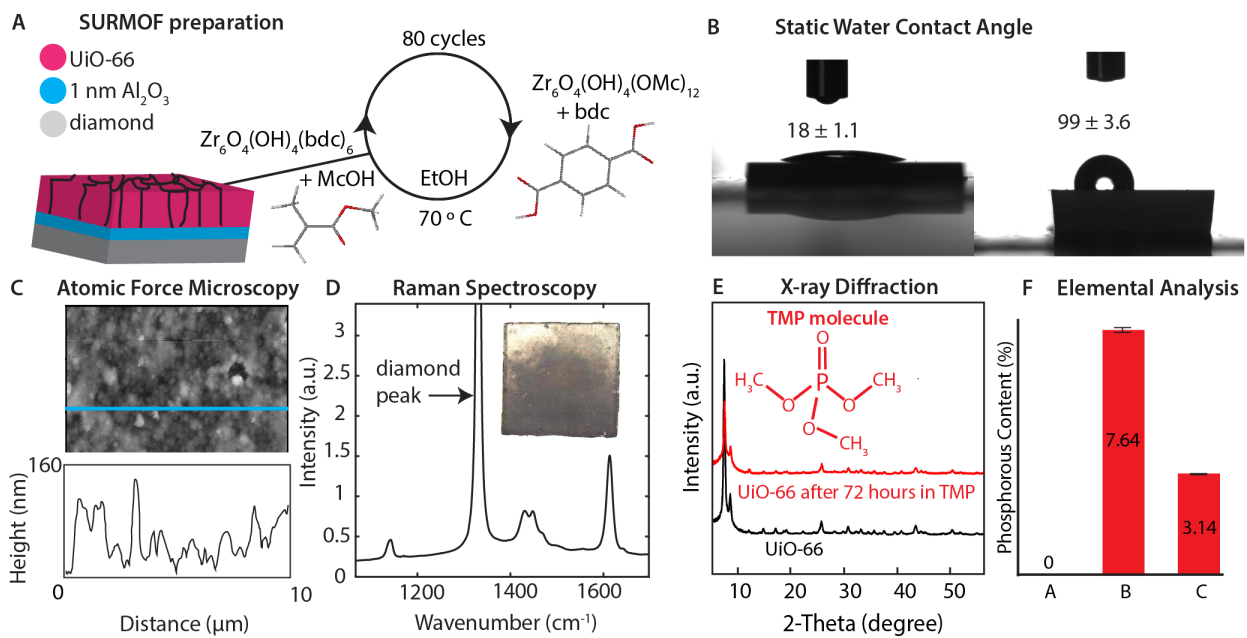

**Figure 2. Characterizing the functionalized diamond surface with UiO-66 SURMOF**. A) Left: Schematic of the functionalized diamond chip with 1 nm of $Al_2O_3$ grown by ALD and ~80 nm of UiO-66 grown by SURMOF (not to scale). Right: chemical process for functionalization of the diamond surface with UiO-66 SURMOF. B) Top: Static water contact angle measurements show the transformation of the hydrophilic $Al_2O_3$-terminated diamond to a hydrophobic surface after the deposition of UiO-66. C) Height profile analysis of atomic force microscopy images yields an average MOF film thickness of 85 nm. D) Raman spectroscopy of the diamond chip functionalized with UiO-66 SURMOF reveals clear modes from the MOF, with peaks at: 1141 $cm^{-1}$ and 1614 $cm^{-1}$, along with a doublet at 1441 $cm^{-1}$. The strong peak at 1332 $cm^{-1}$ corresponds to diamond. Inset: optical microscope image of the functionalized diamond with a size of 2×2 $mm^2$. E) X-ray diffraction of UiO-66 powder still contains distinctive peaks after soaking in trimethyl phosphate for 72 h. Inset: Illustration of the TMP molecule structure. F) Elemental analysis shows no phosphorous in UiO-66 powder (Sample A) but 7.64 ± 0.1 % of phosphorous after soaking in TMP (Sample B), and 3.14 ± 0.01 % after soaking in TMP and washing (Sample C), indicating trapping of TMP within the pores.

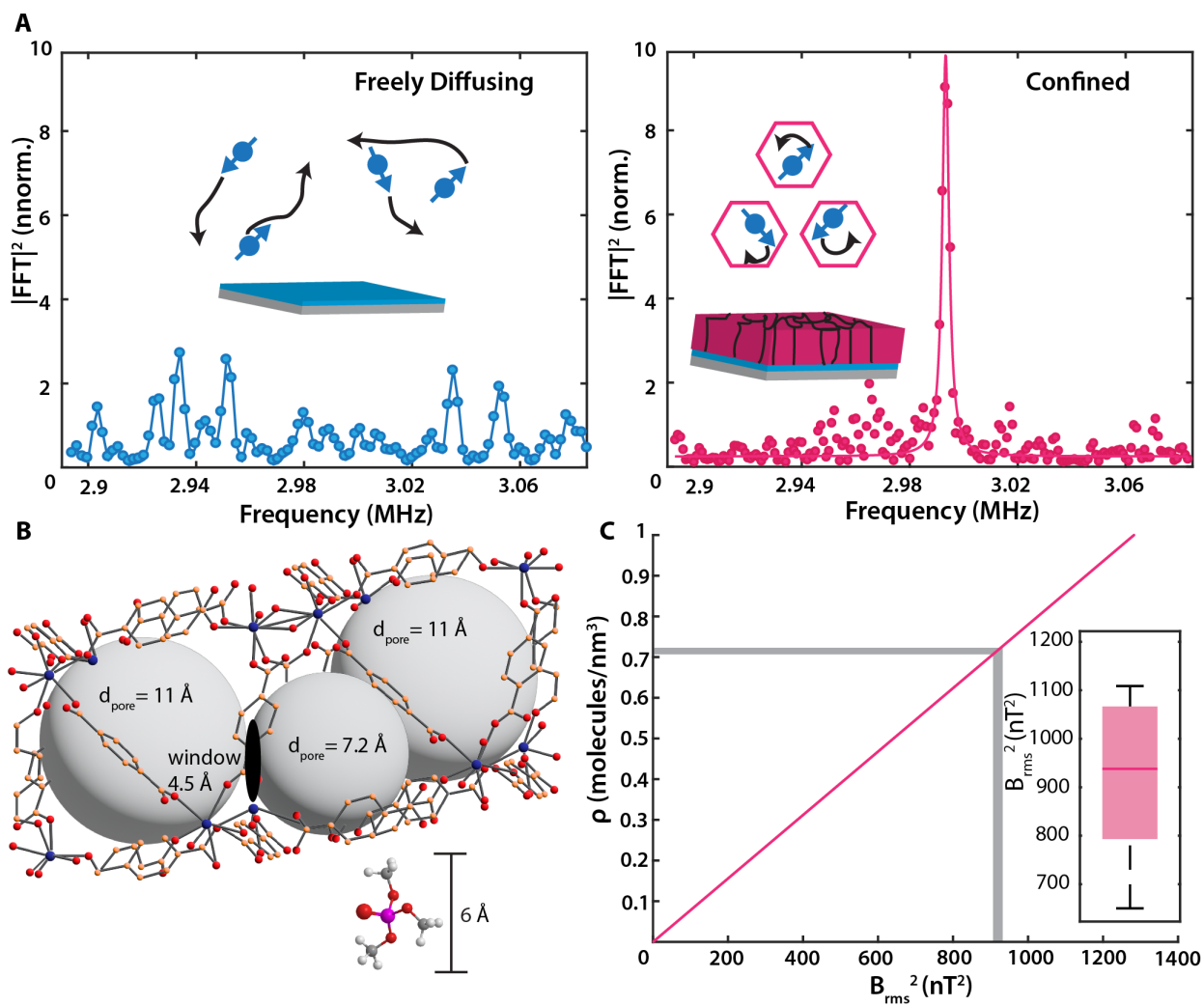

**Figure 3. NV-NMR-signals from small molecules trapped within SURMOF pores.** A) Left: Reference experiment on a $Al_2O_3$ functionalized diamond. No NMR signal can be detected due to fast diffusion of the TMP molecule. Right: The $^{31}P$-NMR signal from TMP measured by NV-NMR shows a narrow signal. Data is normalized to a reference signal. B) UiO-66 with pore sizes shown as spheres and windows connecting the pores. Adapted with permission from {Agrawal, M. et al., J. Phys. Chem. Lett. 10, 7823–7830 (2019)} Copyright {2019} American Chemical Society. A trimethyl phosphate (TMP) molecule with a size of approximately 6 Å can be confined within these pores, restricting its diffusion above the NV-centers. C) The magenta line shows the density of molecules as a function of sensed fluctuating magnetic field ($B_{rms}^2$) and the standard deviation is shown using the thicknesses of the grey lines. Inset: statistics of $^{31}P$ signal strength $B_{rms}^2$ is determined ($n = 4$). The magenta line shows the mean, while the box shows the upper and lower quartiles. Black lines show minimum and maximum values.

## ASSOCIATED CONTENT

**Supporting Information**.

The Supporting Information is available free of charge.

Supporting Information (PDF)

## AUTHOR INFORMATION

**Corresponding Author**

*Email: dominik.bucher@tum.de

**Author Contributions**

§ KSL and XM contributed equally to this work.

DBB conceived the idea of a confined NV-NMR method, designed the experiments and supervised the study. KSL performed SCWA, Raman, NV-NMR experiments and ALD deposition. ALS and XM performed surface modification. XM performed XRD measurements. AH performed AFM measurements. RR, RAF and IDS advised on several aspects of theory and experiments. All authors discussed the results and contributed to the writing of the manuscript.

**Notes**

The authors declare no competing financial interest.

ACKNOWLEDGMENT

This study was funded by the Deutsche Forschungsgemeinschaft (DFG, German Research Foundation) - 412351169 within the Emmy Noether program. AH acknowledges funding from the European Union's Horizon 2020 research and innovation program under the Marie

Sklodowska-Curie Grant Agreement No. 841556. RR acknowledges support from the DFG Walter Benjamin Programme (Project RI 3319/1-1). The authors acknowledge support by the DFG under Germany's Excellence Strategy—EXC 2089/1—390776260 (IDS, RAF and DBB) and the EXC-2111 390814868 (D.B.B.).

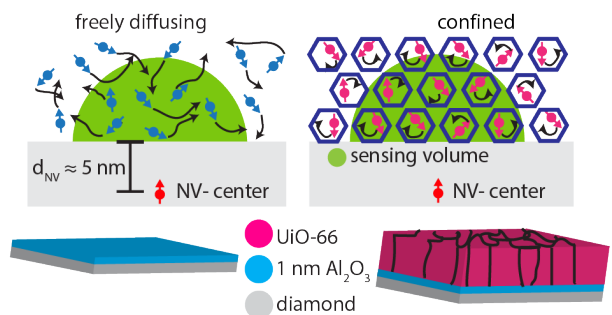

For Table of Contents Only

# Supplementary Information: Using metal-organic frameworks to confine liquid samples for nanoscale NV-NMR

*Kristina S. Liu, Xiaoxin Ma, Roberto Rizzato, A. Lisa Semrau, Alex Henning, Ian D. Sharp, Roland A. Fischer, Dominik B. Bucher*

**This PDF file includes:**

    Elemental Analysis of TMP and UiO-66 MOF powder

    Estimation of pore filling from Elemental Analysis

    Materials and Methods

    Experimental Conditions

    Measuring $^{31}$P signal from dried MOF

    Signal normalization using a 1 MHz reference signal

    Calculation of TMP density within UiO-66 SURMOF pores from Elemental Analysis

**Supplementary Note 1: Elemental analysis of TMP and UiO-66 MOF powder**

The TMP content was determined via elemental analysis, which was performed with Euro EA (HEKAtech). Upon dynamic spontaneous combustion (+Sn=1800°C) and with subsequent chromatographic separation and determination using a thermal conductivity detector, the percent by weight (wt. %) of different elements were determined. The molar ratio between UiO-66 and the guest is approximately 1:6 when the UiO-66 powders were directly filtered without washing after being immersed in TMP liquid for two days, with an experimental P (phosphorous) content of about 7.54 wt.%. This value was obtained by matching the experimental phosphorus content to what is expected from the chemical formulas of UiO-66 and TMP. To remove the TMP adsorbed on the surface, the powders were repeatedly washed. The resulting molar ratio of UiO-66 to TMP was found to be 1:2 with a 3.15 wt.% phosphorous content.

**Table S1.** Elemental analysis of (TMP)x@UiO-66. X in (TMP)x@UiO-66 indicates the amount of TMP based on the best match of the experimental phosphorus value.

| UiO-66 +x Guest(TMP) | C | H | Zr | P |
| --- | --- | --- | --- | --- |
| -- (Experimental) | 25.4 | 2.90 | 24.4 | 3.15 |
| 0 | 34.6 | 1.70 | 32.9 | 0.00 |
| 1 | 34.0 | 2.07 | 30.3 | 1.72 |
| 2 | 33.4 | 2.39 | 28.2 | 3.19 |
| 3 | 32.8 | 2.67 | 26.3 | 4.46 |
| 4 | 32.4 | 2.91 | 24.6 | 5.57 |
| 5 | 32.0 | 3.12 | 23.2 | 6.55 |
| 6 | 31.7 | 3.31 | 21.9 | 7.42 |

**Supplementary Note 2: Estimation of pore filling from elemental analysis**

We estimate the number of pores available to corroborate whether TMP fills the pores of UiO-66 MOF and to estimate to what extent. Our calculations are based on literature porosity values, the spherical geometry of the pores, and the elemental analysis reported above. UiO-66 is characterized by two pores diameters[1], and so an average of the two are calculated to have a volume of ~ $4.5 \times 10^{-28}$ m$^3$. Next, we multiply the literature porosity by the weight of 1 mol of MOF to get the number of pores per mol of MOF. We chose the most suitable porosity considering both literature and commercially available UiO-66 MOF (https://www.strem.com/catalog/v/40-1105/)[2–5].

     0.5 cm$^3$/g × 1664 g = $8.3 \times 10^{-4}$ m$^3$ of pore volume/ mol of MOF.

Then we divide by the average pore volume to get the number of pores:

     $8.3 \times 10^{-4}$ m$^3$ / $4.5 \times 10^{-28}$ m$^3$ ≈ $1.9 \times 10^{24}$ pores/mol of MOF.

Our elemental analysis indicates the presence of 2 mol of TMP/mol of MOF, which would be $1.2 \times 10^{24}$ total TMP molecules. This indicates that approximately 63% of the total pores are filled assuming that each pore can trap 1 TMP molecule.

**Supplementary Note 3: Materials and Methods**

*Diamond preparation.*
An electronic grade diamond purchased from Element Six (1.1% $^{13}$C abundance) was implanted with $^{15}$N (by Innovion, San Jose) and subsequently annealed under vacuum in a home-built oven. The diamond was cleaned before implantation and annealing with a tri-acid cleaning protocol involving equal parts boiling sulfuric, nitric, and perchloric acid according to Brown *et al*.[6]. The implantation was done at an energy of 2.5 keV with an off-axis tilt of 7 degrees and fluence of $2\times10^{12}$/cm². For forming NV-centers, the temperature was ramped up from room temperature to 800 °C with a Tectra BORALECTRIC® sample heater according to the temperature and pressure profile as shown in Figure S1.

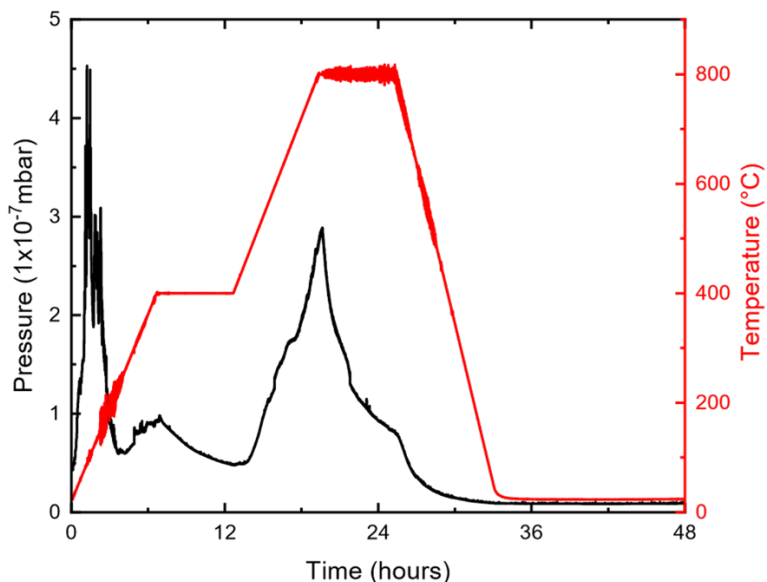

**FIG S1.** Temperature and pressure profile of shallow NV ensemble annealing.

*Atomic layer deposition (ALD).*
Prior to deposition, the diamond surface was cleaned using a piranha cleaning solution comprising a 1:3 ratio of sulfuric acid and hydrogen peroxide for 30 min. The sample was then dried and loaded into the ALD (Veeco Fiji G2 system) chamber. The surface was further cleaned *in-situ* by 5× 0.15 s cycles of ozone exposure, which was produced by electrical discharge in $O_2$ gas. For deposition of $Al_2O_3$ thin films, trimethyl aluminum (98-1955, STREM Chemicals) was used as the aluminum containing precursor and $H_2O$ was used as the oxidant at 200 °C. For deposition of $Al_2O_3$ thin films, trimethyl aluminum (98-1955, STREM Chemicals) as the aluminum containing precursor and $H_2O$ as the oxidant were sequentially introduced into the reactor that was heated to 200° C. This was repeated for 10 cycles to produce a 1 nm thick film of $Al_2O_3$ (with a known nominal growth per cycle of 0.1 nm/cycle)[7]. For reuse of substrates, the $Al_2O_3$ could be removed by soaking overnight in a 5% NaOH solution, after which ALD could again be performed prior to SURMOF deposition.

*Synthesis of cluster-$[Zr_6O_4(OH)_4(OMc)_{12}]$ and fabrication of UiO-66 thin-films (SURMOFs).*
The $[Zr_6O_4(OH)_4(OMc)_{12}]$ (OMc = Methacrylate) was synthesized using the previously described procedure.[8] In detail, 70 Wt.% $Zr(OnPr)_4$ in n-propanol (3.1 mmol, Sigma-Aldrich,) was combined with 1 mL of methacrylic acid (11.8 mmol, McOH, Sigma-Aldrich,) in a flask under Argon at room temperature. After two weeks, the colourless crystals were collected after cleaning with n-propanol (Sigma-Aldrich). The $[Zr_6O_4(OH)_4(OMc)_{12}]$ cluster could be obtained after drying at 80 °C overnight.

The diamond (2×2 mm$^2$) was placed in a double-walled reaction vessel after being exposed to ozone for 5 min. Prior to the experiment, a 0.5 mM $[Zr_6O_4(OH)_4(OMc)_{12}]$ ethanol solution (Solution 1) was mixed with 300 mM McOH, and the deprotonated organic linker solution (Solution 2) was made by combining 3mM terephthalic acid ($H_2BDC$, Sigma-Aldrich,), 1L ethanol (99.9%, Th. Geyer GmbH & Co. KG, Germany), and 320µL trimethylamine ($Et_3N$, Acros). The thin film deposition was carried out through a layer-by-layer (LBL) method in a homemade pump system at 75 °C following a reported procedure[9]. The diamond substrate was initially soaked in Solution 1 for 10 min before being washed in pure ethanol for 2 min. The substrate was then maintained in solution 2 for 10 min before the first synthesis cycle was completed with a 2 min ethanol washing step. The process was repeated 80 times. LabView software was used to control operate of the pump system.

*NV-NMR setup.*
**Diamond alignment and optical approach.** This NV-NMR setup has been previously described in detail in Bucher *et al.* [10]. The diamond was flatly positioned in the center of two neodymium magnets. These magnets were secured to rotation stages, the larger of which (RPO/3(/M), Thorlabs) rotates the magnets to the orientation of the diamond. The smaller rotation stage (RP005(/M)) tilts the magnets in the direction of the NV-centers. This allows for alignment of the $B_0$ field along one of the four possible NV-center orientations. The diamond was glued on one side to a 12 mm thin round glass slide (89015-725, VWR) and a 6 mm glass hemisphere (TECHSPEC® N-BK7 Half-Ball Lenses, Edmund Optics) was glued to the other side. This assembly was glued to a 30 mm cage plate (CP4S, Thorlabs) and fixed onto the experiment 1.2 cm above the top condenser lens (ACL25416U-B, Thorlabs).
**Collection of Photoluminescence.** The photoluminescence (PL) from the NV-centers was collimated by condenser lenses, the bottom one aligned directly above a large area avalanche photodiode (ACUBE-S3000-10, Laser Components GmbH) that was positioned below the diamond. A long-pass filter (Edge Basic 647 Long Wave Pass, Semrock) was placed immediately between the bottom condenser lens and the photodiode to filter out non-NV PL wavelengths. The photo-voltage from the photodiode was digitized with a data acquisition unit (USB-6229 DAQ, National Instruments) and read out by a computer.
**Quantum state initialization and control.** A 532 nm laser (Verdi G5, Coherent) beam was controlled by an acousto-optic modulator (Gooch and Housego, model 3260-220) to pulse the NV-centers with 5 $\mu$s pulses of the laser. The laser power after the AOM was ~ 400 mW when the AOM is fully on. A half-wave plate (AQWP05M-600, Thorlabs) polarized the laser beam in the direction of the NV-centers to maximize contrast. The laser beam was focused onto the diamond using a 125 mm focusing lens (LA1986-A-M, Thorlabs). Microwave frequencies from a signal source (SynthHD, Windfreak Technologies, LLC., New Port Richey) were phase shifted (ZX10Q-2-27-S+, Mini-Circuits) and controlled by switches (ZASWA-2-50dRA+, Mini-Circuits) to generate X and Y pulses. These pulses were then combined (ZX10-2-442-s+) and amplified by a microwave amplifier (ZHL-16W-72+, Mini-Circuits), and delivered to the diamond through a small homemade antenna. The electron spin resonance (ESR) frequency measured from the dip in PL was used to determine the NV resonance frequency at which to perform a Rabi experiment. The Rabi experiment (during which the duration of microwave pulses were swept from 0 to 200 ns) then determined the $\pi/2$ and $\pi$ pulse durations for the correlation spectroscopy pulse sequences. The magnetic field strength $B_0$ can be adjusted by changing the distance between the magnets to tune the nuclei Larmor frequency to that which is suitable for correlation spectroscopy. The correlation spectroscopy sequence has the highest performance for sensing frequencies between ~1-4 MHz because of the $T_2$ relaxation time of the NVs and the finite $\pi$ pulse durations. Therefore, we detect $^{31}P$ (~3 MHz) at 174 mT.

**Emptying and filling SURMOF pores.** The diamond with UiO-66 SURMOF was glued to the assembly previously described above in *NV-NMR Setup*. The glue was cured overnight at 50 °C to ensure stability in trimethyl phosphate. The glued diamond assembly was placed in an oven at 80 °C overnight to empty the ethanol from the pores, after which the diamond was soaked overnight in trimethyl phosphate.

**NV-NMR.** Correlation spectroscopy was performed using XY8-*4* blocks (a total of 32 $\pi$ pulses) with $t_{corr}$ swept starting from 2 $\mu$s to obtain the spectra. For $^{31}P$ detection $t_{corr}$ was swept until 320 $\mu$s in 3201 points. The time domain data were then Fourier transformed and the absolute value plotted using MATLAB. For the $^{31}P$ signal shown in Fig. 3 A, we obtain a SNR of ~12, as calculated by dividing the signal value by the standard deviation of the noise floor within a region without signal. This signal was obtained in 180 min. The spectrum shown is zero-filled with 10000 points (to 1 ms). After zero-filling, the power spectrum (|FFT|$^2$) is plotted, and the linewidth of the resonance is fit with a Lorentzian model[11].

**Determining B$_{rms}$.** The calibration was done according to the protocol described in Liu et al.[12]. A total of 4 different spots were measured on a diamond with SURMOF grown on its surface and TMP filled within the pores, as described above.

**SWCA measurements.** SWCA measurements were performed on an OCA 15Pro contact angle system (DataPhysics Instruments). The acquisition of data and its evaluation were conducted with SWCA 20 - contact angle (DataPhysics Instruments, version 2.0). A 2 $\mu$L droplet of deionized H$_2$O (18.2 M $\Omega$cm at 25 °C, Merck Millipore) was dispensed at a rate of 0.2 $\mu$Ls$^{-1}$ from a 500 $\mu$L Hamilton syringe onto the sample surface. After allowing the droplet to settle for ∼ 3 s, an image was acquired for further processing and quantification of the average Young's contact angle ($\theta\gamma$). This procedure was performed at least 3 times on different spots on the surface, and the standard deviation (error) was determined.

**Atomic Force Microscopy.** *A* MultiMode 8 (Bruker Corp.) was used in tapping mode utilizing NSG30 (TipsNano) to estimate the UiO-66 SURMOF thickness. The thickness of the SURMOF was obtained from the height profiles taken from the mean image data plane of 10 x 10 $\mu$m$^2$ tapping-mode micrographs. Images were analyzed using Gwyddion 2.56.

**Preparation of UiO-66 powder and powder X-ray diffraction (PXRD).** 0.159g ZrCl$_4$ (Alfa Aesar) was added to the 25ml N,N-Dimethylformamide (DMF, Sigma-Aldrich) with stirring at room temperature for 20 min. Then, 0.102g H$_2$BDC was adding to the aforementioned solution. After continuously stirring for 30 min, the homogenous solution was heated in an autoclave at 120 °C for 24 h. After centrifugation and DMF washing, the white powder was soaked in methanol liquid for 3 days for exchange with the DMF in the MOF pores. The UIO-66 powders were collected after centrifugation and dried *in vacuo* at 80 °C. Powder X-ray diffraction (PXRD) patterns were acquired using a Rigaku MiniFlex 600-C diffractometer with a Cu Kα irradiation source (λ = 1.54056 Å) and a scan speed of 10° per min.

**Supplementary Note 4: Experimental Conditions.** The properties of our shallow NV-ensemble in diamond after the growth of a UiO-66 SURMOF layer on the surface are depicted in Table S2. We used the NV resonance for the $m_s = 0$ to -1 transition to determine the magnetic field strength. The contrast of the Rabi oscillations is on average ~ 4.2 %. The spin-lattice relaxation time $T_1$ was fit to an exponential function ($2.2 \pm 0.79$ ms). The spin-spin relaxation time $T_2$ was fit to a stretched exponential function ($5.8 \pm 0.35$ μs).

**Table S2:** NV ensemble Rabi contrast, relaxation and coherence properties

| Contrast (%) | $T_1$ (ms) | $T_2$ (μs) |
|---|---|---|
| $4.2 \pm 0.83$ | $2.2 \pm 0.79$ | $5.8 \pm 0.35$ |

**Supplementary Note 5: Measuring $^{31}$P signal from dried MOF.**

To ensure that the detected signal comes from within the pores of the MOF, after filling the pores with TMP the liquid was removed from the assembly and was rinsed with $H_2O$, in which TMP is soluble. Then the MOF-coated diamond was soaked in $H_2O$ for 1 h and dried with nitrogen. A strong $^{31}$P signal is still present, indicating that the signal originates from TMP within the pores since there are no other sources of $^{31}$P.

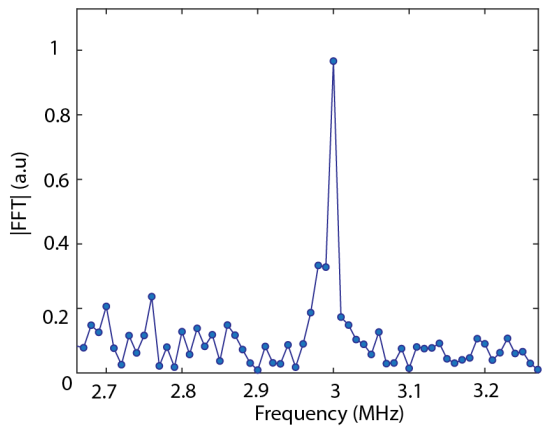

**FIG. S2.** Left: $^{31}$P signal from MOF rinsed and soaked in $H_2O$ and then dried with nitrogen indicates that the signal originates from within the pores of the MOF.

**Supplementary Note 6: Signal normalization using a 1 MHz reference signal.**

To accurately compare between the $^{31}$P signal from UiO-66 SURMOF soaked in TMP and $Al_2O_3$ soaked in TMP we use a 1 MHz reference signal (see Material and Methods: Determining $B_{rms}$). Variations in the resulting signal size from a signal source of the same strength would be due to differences in experimental conditions. To ensure that the lack of signal in Figure 3 A of the main paper is not due to much less sensitive experimental conditions, we normalize the data using correlation spectroscopy data of the 1 MHz signal measured to 80 μs in 401 points obtained before each $^{31}$P correlation experiment. Due to differences between experiments such as contrast and coherence properties, the $^{31}$P experiment on only $Al_2O_3$ is 67% as sensitive as the $^{31}$P experiment with the SURMOF.

**Supplementary Note 7: Calculation of TMP density within UiO-66 SURMOF pores from Elemental Analysis.**

From the molar ratio obtained from elemental analysis of UiO-66 MOF powder, we can make an estimate for the density of TMP molecules to corroborate our calibration and whether the NV-NMR signal size agrees with the porosity of UiO-66. In a prior publication of the UiO-66 SURMOF growth method[9] it was found that 9700 ng of UiO-66 SURMOF was deposited on a 14 mm diameter quartz crystal microbalance (154 mm$^2$ circular area). In our laser spot of 4000 μm$^2$, that would correspond to 0.25 ng or $1.5\times10^{-13}$ mol of UiO-66 and $3\times10^{-13}$ mol of TMP. Multiplying by Avogadro's number gives $1.8\times10^{11}$ molecules of TMP. Dividing this by the total cylindrical volume of $3.4\times10^{11}$ nm$^3$ (calculated from the area of the laser spot multiplied by 85 nm average height of the UiO-66 SURMOF) yields a density of ~ 530 molecules of TMP/ (10nm)$^3$. This agrees well with the NV-NMR experimental calibration.